\documentclass[preprint,aps2]{revtex4}

\usepackage{graphicx}

\newcommand{\icrn}{$I_{\mathrm{c}}R_{\mathrm{n}}$}
\newcommand{\jcrn}{$J_{\mathrm{c}}\rho_{\mathrm{n}}$}
\newcommand{\critical}[1]{$#1_{\mathrm{c}}$}
\newcommand{\Tc}{\critical{T} {}}

\newcommand{\jc}{\critical{J}}

\newcommand{\jcgb} {$J_ \mathrm{c}$}

\newcommand{\etal}{{\sl et\,al.}}
\newcommand{\degree}{$^{\circ}$}

\newcommand{\rhon}{$\rho_{\mathrm{n}}$}

\begin{document}

\title{\,
\vspace*{8 mm}
{}Properties of Grain Boundaries in High-\Tc Superconductors
\newline
-- Notes on a Recent Presentation  --}

\author{{}
\vspace*{2cm}
J. Mannhart}

\affiliation{{}
\vspace*{2cm}
Experimentalphysik VI, Center for Electronic Correlations and Magnetism,
Institut f\"{u}r Physik, Universit\"{a}t
Augsburg, D-86135 Augsburg, Germany \vspace{4cm}}


\begin{abstract}
The purpose  of this article is to discuss a view concerning key datasets of the properties of
grain boundaries in high-\Tc superconductors  that was recently expressed in Ref.\,\cite{GPC}. 
The reference also criticizes our research.  Using examples I disprove this criticism. \end{abstract}

\vspace{1cm}

\maketitle


Grain boundaries in high-\Tc superconductors are a topic of intense
interest. Their low critical current densities provide a severe obstacle
for the realization of superconducting wires that operate at 77\,K. 
It is therefore important that the critical 
currents can  be significantly 
enhanced if 
the  grains are aligned, as reported by my colleagues and myself, for example in \cite{Dimos1988, Dimos1990}. 
Our finding provides the basis for the development of state-of-the-art high-\Tc wires, 
the so-called coated conductors. This field is very active, and many groups have made outstanding contributions.\\ 

In the following I will refer to the publications \cite{Dimos1988, Dimos1990} as ``ours". The author of  \cite{GPC} did not contribute to our studies. \\

The recent Ref.\,\cite{GPC} comments on our work, and in doing so caused misunderstandings in the community.
Aiming to provide clarifications, I will address two problems: 
First, in \cite{GPC}  its author admits  that over many years he published our results without proper references, such as if they had been obtained by him or by his group (\cite{GPC},  p.\,107,109).  While he identifies  only in exceptional cases the figures in which he displays our results, he recognizes that  by publishing our data without naming their source he did not meet  scientific standards  (\cite{GPC}, p.\,109). However, it also seems that he has misinterpreted some of our data. 
Second, Ref.\,\cite{GPC} contains a number of invalid or misleading statements.  \\

The following discussion refers only to part of the work of the author of \cite{GPC}.
He contributed to the progress of the field with very good studies that are
not subject to the problems discussed here. \\

His coauthors
had little
opportunity to note the problems in their papers, since their
articles usually refer to his publications  as sources of the data under
concern. 
To the coauthors, the citation of these papers, which appear
to correctly show original measurements, must have seemed
perfectly appropriate. \\

\vspace{0.4cm}
\newpage
\noindent
\textbf{1. Analysis of our Data}\\

As I will illustrate in the following with an example,  it is not always possible for me to reproduce the analysis of our data published by the author of \cite{GPC}.\\

In the community it is  controversially debated whether the  
boundaries'
critical current densities \jc{} and normal state resistivities \rhon{} scale, for example like $J_{\mathrm{c}}\rho_{\mathrm{n}}  
\propto {1/ \rho ^ {3/2} _{\mathrm{n}}}$, 
as argued by him. This scaling is predicted by a model of the grain boundary mechanism, the intrinsically-shunted-junction (ISJ) model which he supports.\\

Figure\,1 presents  the \jcrn{} products of bicrystals plotted as a function of \jc{} 
as published by him in \cite{GPC}, where he explains  that the black datapoints of Fig.\,1 are our data of table\,I \cite{Dimos1990}.  I agree that several of these are our data: their \jc{} values, their  \jcrn{} products, and thereby also their  \rhon{} values are those that we have measured and published. Furthermore, other data of Fig.\,1  ($7.9\times10^4 \mathrm{A/cm}^2$, 3.75\,mV; $4.7\times10^5 \mathrm{A/cm}^2$, 5.4\,mV) that are not identified by black dots as ours, are the same as remaining data of table\,I \cite{Dimos1990}.\\

If the data are plotted as a function of 1/\rhon{}, do they follow the proclaimed scaling behavior?  According to the author of \cite{GPC}, our data scale well (Fig.\,2) and thus provide evidence  for the ISJ-model. 
However, it is not possible for me to reproduce the presentation of our data he is showing. The data  that we measured and published are in clear disagreement with the scaling rule (Fig.\,3) \cite{measurement errors}. Why is it, that in his figures our data follow the scaling?  \\

In Ref.\,\cite{GPC} p.\,108 its author mentions for the first time that he had "reevaluated" our data of Fig.\,3, selecting some and disregarding other data. He did not contribute to our studies, but he is of course welcome to use the data. Because we have evaluated our results correctly, I do not, however, see a possibility to improve the evaluation, and I am interested to know how he tried to do so.  Unfortunately, \cite{GPC} is very brief in this respect.  \\

As a comparison of Fig.\,2 with Fig.\,3 shows, he furthermore apparently misinterpreted our data. To a large  extent, the agreement with his model seems to originate from modifications of the  \rhon{} values of our data published in table\,I \cite{Dimos1990}.  Because the apparent modifications are extensive and change the conclusions to be drawn from our measurements, it is important to know how he derived the specific values of his datapoints (Fig.\,2) from our measurements (Fig.\,3).  It is also important to understand why he still uses for his \jcrn{}(\jc{}) plots our \rhon{} values with little changes only (see Fig.\,1 of \cite{GPC}). His publications do not provide answers to these questions. Refs.\,\cite{Gross1991, Gross1994}, for example, present the data with the altered values, but do not apprise the reader of the data's origin or the interpretation they were subjected to.\\

Ref.\,\cite{GPC} does not clarify  the modifications of our data either, but criticizes  our publication \cite{Hilgenkamp2002} for not stating that the author of \cite{GPC} found the scaling behavior  at approximately the same time as another group \cite{Russek} did (see the comment presented as Ref.\,17  in \cite{GPC}). As we describe in \cite{Hilgenkamp2002}  on  p.\,506, Ref.\,\cite{Russek} provided evidence that the  \icrn{} products of 45\degree{} boundaries scale with  \jc{} and \rhon{}, in particular  the \icrn{} products of individual boundaries that are gradually depleted from oxygen.  R.\,Gross proposes a more universal  scaling: he expects the  \icrn{} products to scale with \jc{} and \rhon{} if many boundaries with a variety of boundary angles are compared. However, he has not provided convincing evidence   that  this universal scaling  exists. As discussed above, the datasets  he presented as evidence are, for example, not consistent with each other. His complaint is therefore unwarranted.\\

\vspace{0.4cm} 
\noindent
\textbf{2. Comments on our Work Made in \cite{GPC}}\\

A large number of statements on our work presented in Ref.\,\cite{GPC} are invalid or misleading, as I will illustrate with a second example: \\

Fig.\,4 presents data on the angular dependence of the grain boundary critical current density \jc{},
plotted such that the datasets can be easily compared. It is evident
that many  data of R.\,Gross shown in  b) and d) are the same
as  the data of us and Z.\,Ivanov \etal{}  shown in  a) and c), respectively. However, 
Refs.\,\cite{Gross1991,Gross1994,Gross1994b}, in which b), d)  
or closely related figures were published, do not use appropriate citations. They present the figures as if all data were his.\\

My coauthor and I therefore cautioned in 
\cite{Hilgenkamp2002} {} (p.\,498):
 {\it ``Data showing an
exponential   \jc($\theta$) dependence and critical current densities
of [100]-tilt and [100]-twist boundaries have also been published 
by Gross and Mayer (1991) \cite{Gross1991} and Gross (1994a) \cite{Gross1994}. 
It seems that many of these data
are reproductions of data given by Dimos 
\etal{\,}(1990) \cite{Dimos1990} and Ivanov \etal{\,}(1991) \cite{Ivanov1991}.''} The 
numbering of the references was
adapted to fit the present manuscript. 
As Fig.\,4 shows, our comment is correct and points towards a serious problem.

Instead of resolving this problem,  the author of \cite{GPC} lists in the Ref.\,43 of    \cite{GPC} 
our statement in   
an altered form. He ignores that  in addition to his paper   {\it Gross and Mayer (1991) \cite{Gross1991} } our statement also concerns his article  {\it Gross (1994a) \cite{Gross1994}}.  
Although  modified by him only slightly, the statement now conveys an incorrect message, 
against which he is then arguing: \\

{\it ``In the recent review \cite{Hilgenkamp2002} it is stated with respect Ref.\,\cite{Gross1991}: It
seems that large part of the data are reproductions of data
given by Dimos \etal{\,}1990 (\cite{Dimos1990})  and Ivanov \etal{\,}1991 (\cite{Ivanov1991}). This is
indeed true with respect to the data of Dimos \etal{\,\,}that
have been included by courtesy of Dimos and Chaudhari.
However, this is not true with respect to data of Ivanov
\etal{} The paper by Ivanov \etal{\,}\cite{Ivanov1991} has been submitted about half a year later than the paper by Gross and Mayer \cite{Gross1991}. The data by Ivanov \etal{}{\,\,}have been included in a subsequent review \cite{Koelle} by courtesy of Ivanov \etal"} (cited from Ref.\,43 of \cite{GPC}). The 
numbering of the references was
adapted to fit the present manuscript. \\

The sequence of the submission and publication dates is as follows: The data of Dimos \etal{}\,\,were published first by Dimos \etal{}\,\,in \cite{Dimos1990} (submitted 1989, published 1990) and afterwards by the author of \cite{GPC},  beginning with \cite{Gross1991} (published in 1991  without note to the submission date). The data of Ivanov \etal{}\,\,were published first by Ivanov \etal{}\,\,in \cite{Ivanov1991}  (submitted and  published 1991). The author of \cite{GPC} published afterwards  the data of Fig.\,4d in \cite{Gross1994, Gross1994b}, which both appeared without published submission date in 1994.\\

Although neither 
 \cite{Gross1994, Gross1994b},  nor \cite{GPC} state so,
it has to be concluded that the data of  \cite{Gross1994, Gross1994b} shown in Fig.\,4d 
are the data of  Ivanov \etal{\,}\cite{Ivanov1991}. Our cautionary comment made in  \cite{Hilgenkamp2002} {}on p.\,498 is therefore correct.\\

In summary,  a large number of comments on our work presented in \cite{GPC} are invalid. I disproved some of them. 
By  correcting, where possible,  his articles in which he incorrectly published our results or those of others, the author of  \cite{GPC} can resolve and settle the problems he started to address in \cite{GPC}.\\

I am grateful to all who
supported me in writing this note.

\newpage
\clearpage

\noindent
\textbf{References}
{}

\newpage
\clearpage

\noindent
\textbf{Figure Legends}\\

\noindent

\noindent
\textbf{Fig.\,1}\\
Data from an article by R.\,Gross, showing \jcrn {} products of bicrystal grain
boundaries  
plotted as a function of the grain boundary critical current density \jc.
As is evident from \cite{Gross1990} and as he explains in \cite{GPC}, the solid data are our data from  table\,I of \cite{Dimos1990}. \\
\noindent
The line shows the  scaling behavior predicted by the ISJ-model. \\
\noindent
The data and this line are from \cite{Gross1990}, Fig.\,3.\\

\noindent
\textbf{Fig.\,2}\\
Similar dataset as Fig.\,1, taken from another publication of R.\,Gross  \cite{Gross1991}. The  \jcrn {} products are  plotted here  as a function of the inverse boundary resistivity 1/\rhon. The match between this dataset and the dataset of Fig.\,1 proves that both were obtained from  the same samples.  The solid data are therefore our data  published in table\,I of \cite{Dimos1990}.\\
\noindent
The line shows the expected scaling behavior  predicted by the ISJ-model. In this plot, our data follow the scaling behavior. \\
\noindent
The data and this line are from  \cite{Gross1991}, Fig.\,3a  (see also 
\cite{Gross1994}, Fig.\,6.5). \\

\noindent
\textbf{Fig.\,3}\\
The  \jcrn {} products  of our samples as measured and published (\cite{Dimos1990},  table\,I), plotted as a function of 1/\rhon.
Because the seven solid data of Fig.\,2 are from the same dataset (\cite{Dimos1990},  table\,I),  they have to be identical to seven of the datapoints shown here, which is not the case. \\
\noindent
The line shows the scaling behavior 
according to Fig.\,3a of \cite{Gross1991}. 
Our data do not scale. \\
\noindent
One datapoint of  table\,I, \cite{Dimos1990}
(\icrn=0.06\,mV, \rhon=0.02$\,\Omega \mu\textrm{m} ^2$) 
has not been plotted (see the comment given in \cite{Gross1990} as Ref.\,14). \\

\newpage
\noindent
\textbf{Fig.\,4}\\
Grain
boundary critical current densities \jc{} measured as a function 
of the  boundary angle.\\

\noindent
a) Our data on the angular dependence of \jc{}, taken from table\,II of \cite{Dimos1990}.\\

\noindent
b) Data on the angular dependence of \jc{}  taken from  R.\,Gross \etal{\,\,}(Fig.\,2 of \cite{Gross1991}). Together with
additional data, these data were also published  in  Fig.\,6.4a of \cite{Gross1994} and in
Fig.\,3a of \cite{Gross1994b}, always as if all data were his.  A comparison with a) shows, however, that part of the data result from our work \cite{Dimos1990}, which is now confirmed by \cite{GPC}.\\

\noindent
c) Data from Z.\,Ivanov \etal{} (Fig.\,4 of \cite{Ivanov1991}), showing the \jc{} of their samples measured as a function 
of the  boundary angle.\\

\noindent
d) Data from R.\,Gross (Fig.\,6.4b of \cite{Gross1994}, 
also published in Fig.\,3b of \cite{Gross1994b}), showing again \jc{}  as function of the angle.  In these publications the figure is presented as if all data were data of his.  It seems, however, that they are reproductions of the data of  Z.\,Ivanov \etal{} shown in c).\\

\noindent
The data shown in a) and b) 
have been measured at 4.2\,K, those of c) and d) at 77\,K. 
The hatched lines depict  exponential \jcgb($\theta$)-dependences.
In a) one datapoint with $ \Theta $~=~4\degree {} and $ \Phi $~=~4\degree {} 
has not been plotted, 
because this boundary has two equally strong tilt-components.

     {~}
      \begin{figure}
      \hspace{-4cm}
      \includegraphics[width=20cm]{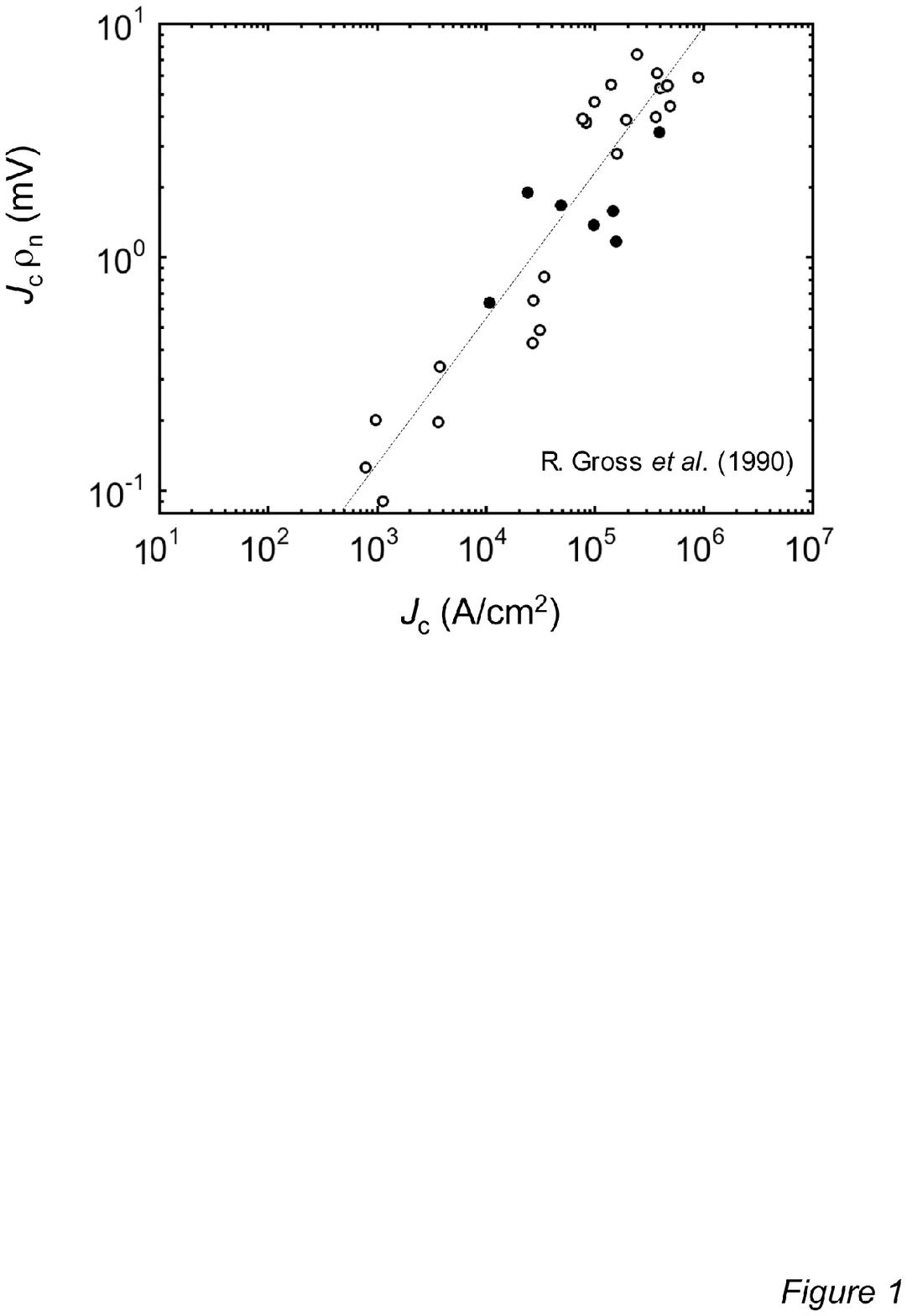}
      \end{figure}
      {~}
      \begin{figure}
      \hspace{-4cm}
      \includegraphics[width=20cm]{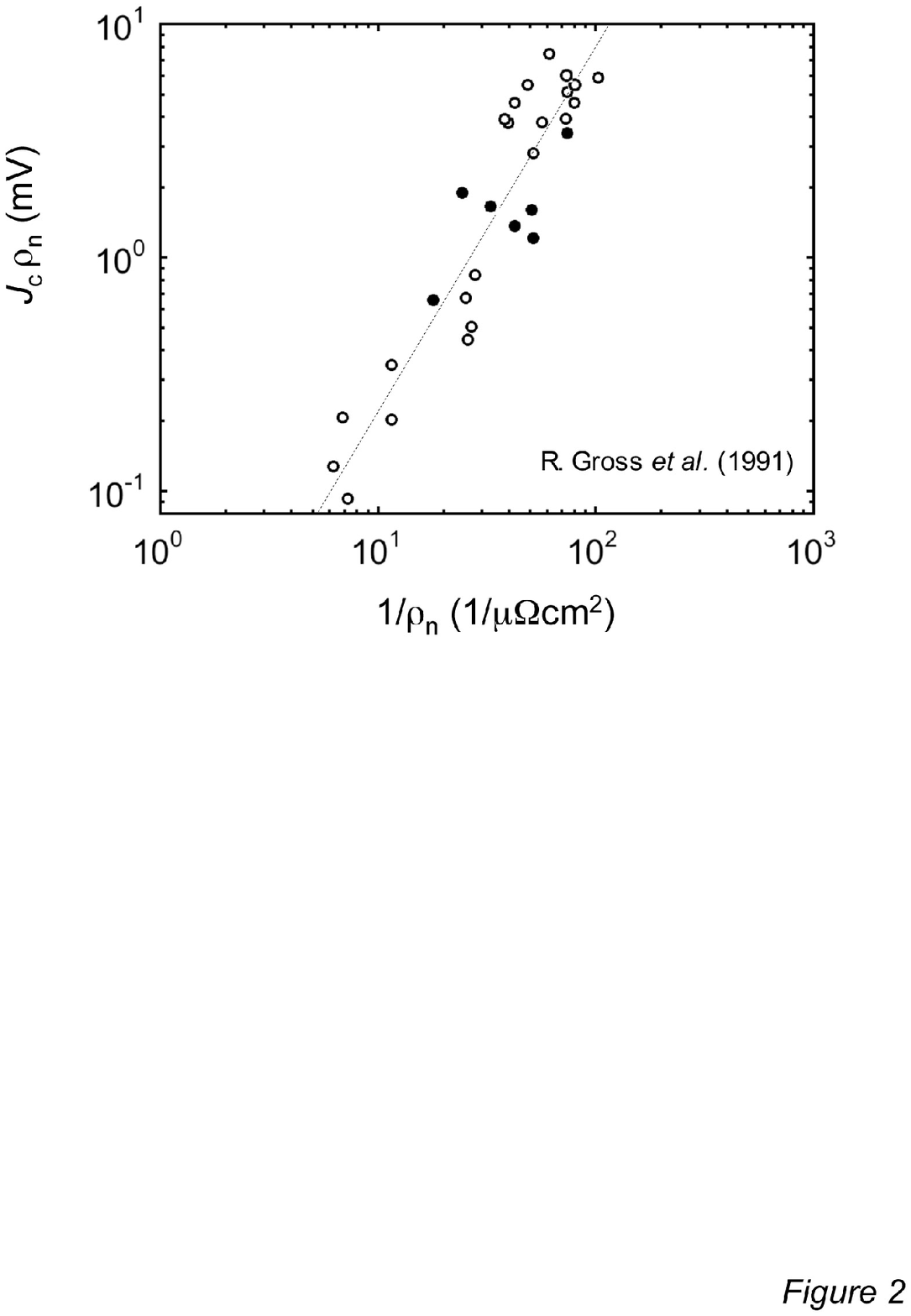}
      \end{figure}
     {~}
      \begin{figure}
      \hspace{-4cm}
      \includegraphics[width=20cm]{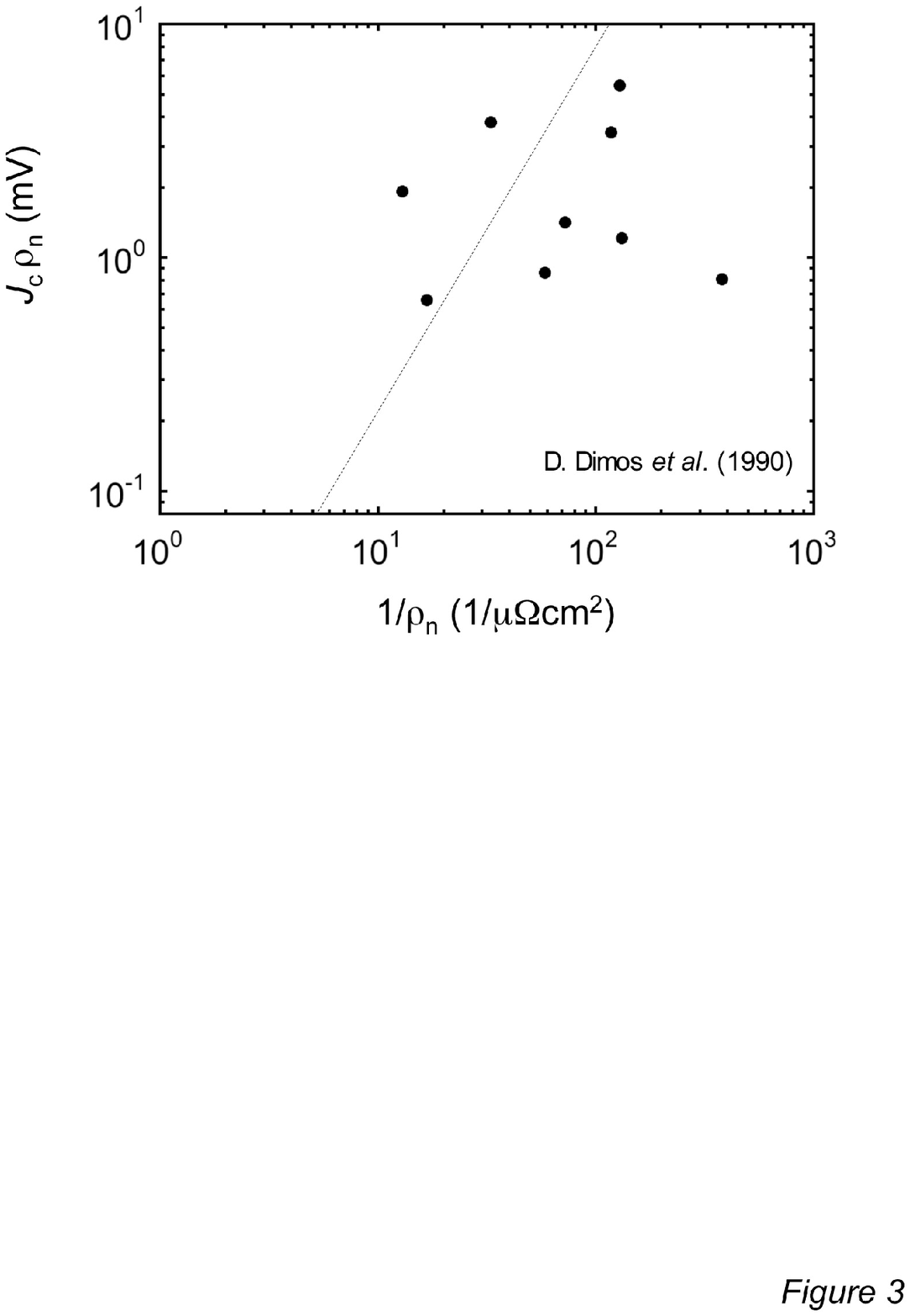}
      \end{figure}
     {~}
      \begin{figure}
      \hspace{-4cm}
      \includegraphics[width=20cm]{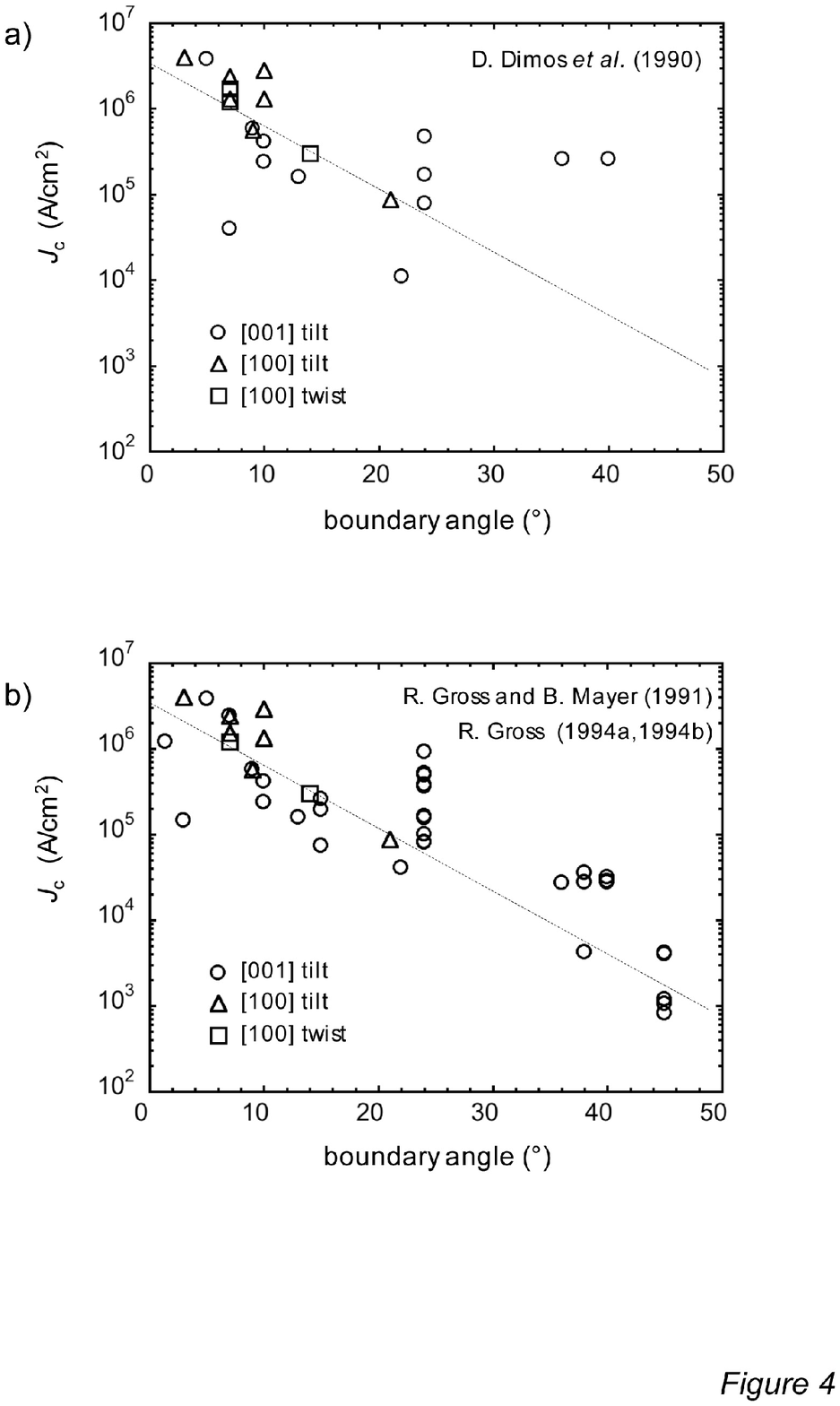}
      \end{figure}
     {~}
      \begin{figure}
      \hspace{-4cm}
      \includegraphics[width=20cm]{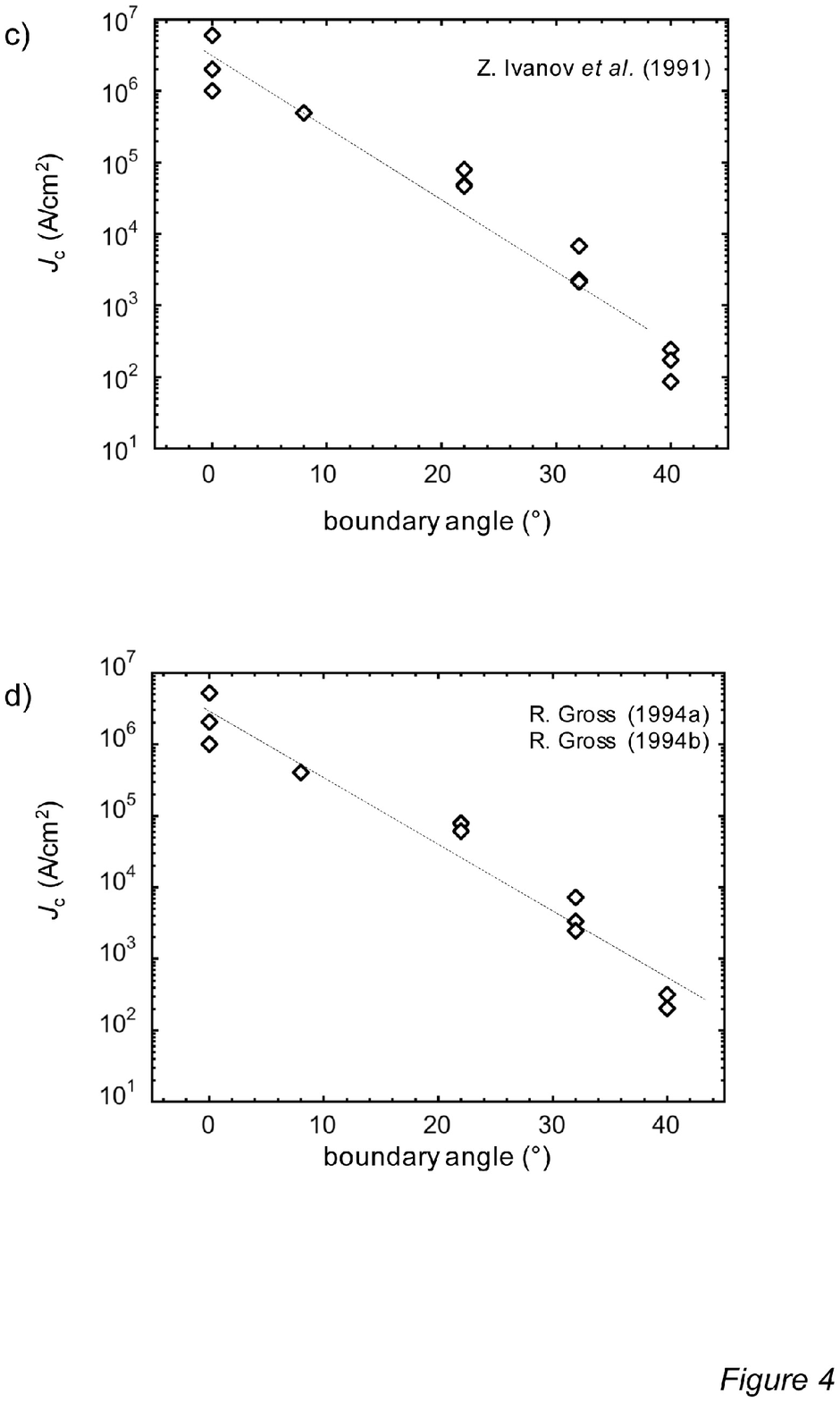}
      \end{figure}

\end{document}